\documentclass[conference]{IEEEtran}

\usepackage[stretch=30]{microtype}
\usepackage[font=small,labelfont=bf]{caption}

\makeatletter
\usepackage{scrpage2,enumitem,float,graphicx,setspace,listings,comment,multicol,caption,subcaption,siunitx,stfloats,acronym,tabularx,multirow}
\usepackage[utf8]{inputenc}

\usepackage[backend=biber,style=ieee,sorting=none,giveninits=true]{biblatex}
\usepackage[usenames,dvipsnames]{xcolor}
\usepackage{todonotes}
\usepackage{tikz}
\usepackage{textcomp}
\usepackage{hyperref}
\usepackage{xcolor}
\hypersetup{
    colorlinks,
    linkcolor={red!50!black},
    citecolor={blue!50!black},
    urlcolor={blue!80!black}
}
\usepackage{lipsum}
\newcommand\copyrighttext{%
  \footnotesize \textcopyright 2018 IEEE. Personal use of this material is permitted.
  Permission from IEEE must be obtained for all other uses, in any current or future
  media, including reprinting/republishing this material for advertising or promotional
  purposes, creating new collective works, for resale or redistribution to servers or
  lists, or reuse of any copyrighted component of this work in other works.
}
\newcommand\copyrightnotice{%
	\begin{tikzpicture}[remember picture,overlay]
	\node[anchor=south,yshift=10pt] at (current page.south) {\fbox{\parbox{\dimexpr\textwidth-\fboxsep-\fboxrule\relax}{\copyrighttext}}};
	\end{tikzpicture}%
}

\lstdefinelanguage{p4}
{
	keywords={if,else},
	keywords=[2]{apply,valid,select,current,extract,add_header,copy_header,remove_header,modify_field,add_to_field,add,set_field_to_hash_index,truncate,drop,no_op,push,pop,count,meter,generate_digest,resubmit,recirculate,clone_ingress_pkt_to_ingress,clone_egress_pkt_to_ingress,clone_ingress_pkt_to_egress,clone_egress_pkt_to_egress,register_write,register_read},
	keywords=[3]{\#include},
	keywords=[4]{length,fields,max_length,width,layout,attributes,type,static,result,direct,instance_count,min_width,saturating,exact,ternary,lpm,range,valid,mask},
	keywords=[5]{bytes,packets,control,action,table,counter,header_type,header,register,parser,metadata,primitive_action,meter,parse_error,default},
	keywords=[6]{reads,actions,min_size,max_size,size,support_timeout,action_profile},
	keywordstyle=\bf,
	keywordstyle=[2]\bf,
	keywordstyle=[3]\bf,
	keywordstyle=[4]\bf,
	keywordstyle=[5]\bf,
	keywordstyle=[6]\bf,
	sensitive=true, 
	morecomment=[l]{//},
	morecomment=[s]{/*}{*/},
	commentstyle=\em,
	morestring=[b]"
}

\author{\IEEEauthorblockN{Jonathan Vestin}
\IEEEauthorblockA{Dept. of Maths and Computer Science\\
Karlstad University\\
65188 Karlstad, Sweden\\
Email: jonathan.vestin@kau.se}
\and
\IEEEauthorblockN{Andreas Kassler}
\IEEEauthorblockA{Dept. of Maths and Computer Science\\
Karlstad University\\
65188 Karlstad, Sweden\\
Email: andreas.kassler@kau.se}
\and
\IEEEauthorblockN{Johan {\AA}kerberg}
\IEEEauthorblockA{ABB Corporate Research\\
V\"{a}steras, Sweden\\
Email: johan.akerberg@se.abb.com}}

\makeatother

\newcommand{\code}[1]{\texttt{#1}}
\newcommand{\ngfig}[4]{
	\begin{figure}
    \centering
	\includegraphics[width=#4]{#1}

	\caption{#3}

	\label{fig:#2}
	\vspace{-1ex}
	\end{figure}
}
\newcommand{\nfig}[3]{
	\ngfig{#1}{#2}{#3}{1\columnwidth}
}

\newcommand{\note}[1]{
}

\newcommand{\tododone}[1]{}

\newacro{iot}[IoT]{Internet of Things}
\newacro{plr}[PLR]{Packet Loss Rate}
\newacro{rtt}[RTT]{Round Trip Time}
\newacro{bfd}[BFD]{Bidirectional Forwarding Detection}
\newacro{netem}[netem]{Network Emulator}
\newacro{tbf}[TBF]{Token Bucket Filter}
\newacro{ovs}[OVS]{Open vSwitch}
\newacro{sdn}[SDN]{Software Defined Networking}
\newacro{netns}[netns]{Network Namespaces}
\newacro{lxc}[LXC]{Linux Containers}
\newacro{loess}[LOESS]{Locally Weighted Scatterplot Smoothing}
\newacro{lm}[LM]{Linear Model}
\newacro{icn}[ICN]{Industrial Control Network}
\newacro{p4}[P4]{Programming Protocol-Independent Packet Processors}
\newacro{prp}[PRP]{Parallel Redundancy Protocol}
\newacro{mrp}[MRP]{Media Redundancy Protocol}
\newacro{nos}[NOS]{Network Operating System}
\newacro{cnf}[CNF]{Conjunctive Normal Form}

\begin{document}
\title{FastReact: In-Network Control and Caching for Industrial Control Networks using Programmable Data Planes}

\maketitle
\copyrightnotice

\begin{abstract}
Providing network reliability as well as low and predictable latency is important especially for Industrial Automation and Control Networks. However, diagnosing link status from the control plane has high latency and overhead. In addition, the communication with the industrial  controller may impose additional network latency. We present \emph{FastReact} - a system enabling In-Network monitoring, control and caching for Industrial Automation and Control Networks. \emph{FastReact} outsources simple monitoring and control actions to evolving programmable data planes using the P4 language. As instructed by the Industrial Controller through a Northbound API, the SDN controller composes control actions using Boolean Logic which are then installed in the data plane. The data plane parses and caches sensor values and performs simple calculations on them which are connected to fast control actions that are executed locally.  
For resiliency, \emph{FastReact} monitors liveness and response of sensors/actuators and performs a fast local link repair in the data plane if a link failure is detected. Our testbed measurement show that \emph{FastReact} can reduce the sensor/actuator delay while being resilient against several failure events.

\end{abstract}
\acresetall

\section{Introduction}
\label{sec:intro}
The domain of industrial automation and control has gained a lot of attention recently through the introduction of \ac{iot}, including smart cities, home automation, self-driving cars, wearable IT and industry 4.0. These new use cases put great pressure on manufacturers to perform fast innovation, where each customer expects individually customized solutions. \acp{icn} must cope with such increased expectations, and allow for cloud-enabled re-configurable and re-programmable networking. However, at the core of automation and control is essentially the reliable and timely delivery and exchange of data. Consequently, the systems need to be designed for   high availability and predictable low latency even under extreme circumstances, such as during link or equipment failures. This introduces significant challenges in designing a network which connects sensors, controllers and actuators at low latency in a flexible way. 

In todays industrial setup, the integration between network and control is quite simple and the control is highly centralized and customized to the use case. Typically, the sensors generate (periodic) data, which are transferred to an industrial controller or an intermediate proxy node that analyzes the data. Based on the use case and received sensor values, the controller derives control actions and sends them to actuators. Also, the controller can modify network paths, link configurations or proxy locations. Nevertheless, the communication to the industrial controller may involve many hops, each one prone to failure and additional latency. In addition, the network protocols and processing of data packets in the switch are decoupled from the application logic, which consequently requires end-to-end control logic, involving actuators/sensors/controllers. 

Recently, \ac{sdn} has shown to be an interesting architectural solution to make networks more flexible and programmable. In \ac{sdn}, the control and data plane are decoupled. The control plane is opened up and logically centralized using a so called SDN controller, which enables faster innovation, manufacturer independent configuration and rapid deployment of custom network functions. A \ac{nos} continuously monitors and reconfigures the network according to customer needs, and provides monitoring statistics to find deployment issues and opportunities for service improvement. Industrial automation has seen an increasing interest in utilizing the advantages which \ac{sdn} provides, but there still remains concern for unpredictable latency and reduced reliability. However, proactive failure recovery (e.g. fast failover~\cite{web:openflow13-spec}) and redundant controllers (e.g. ONOS~\cite{berde14}) provide solutions for enabling high availability in \ac{sdn}-based industrial networks. Especially the recent development of programmable data planes using e.g. P4 ~\cite{bosshart14} opens up previously unexplored use cases for industrial automation and control networks, removing many limitations of the southbound protocol OpenFlow~\cite{tech:openflow}. 


This paper presents \emph{FastReact}, which advocates the idea to outsource parts of the industrial controller logic to the data plane of new emerging programmable switch architectures. \emph{FastReact} provides in-network sensor monitoring, control and data caching for industrial automation and control networks. In \emph{FastReact}, programmable switches parse sensor packets at line rate and cache a history of sensor values in custom data structures in the switch. Highly customizable  control logic is run inside the data plane and the switch can trigger local control actions from the data plane. The logic is specified through conditionals composed of boolean logic and comparison functions, which are installed by the SDN controller, as instructed from the Industrial Controller for the given use case. \emph{FastReact} enables rapid in-network decision-making based on observed sensor data, historical sensor data, or preconfigured lightweight statistical functions, such as moving averages. Partially relocating the control logic to the data plane significantly reduces the path between sensors and actuators, improving latencies while also reducing potential points of failure. Furthermore, as high availability is critical in \ac{icn}, \emph{FastReact} supports resiliency through link liveness monitoring, performing proactive link repair if a link or piece of equipment fail. We have implemented \emph{FastReact} in the P4 language, and evaluated it using the CORE network emulator. We find that \emph{FastReact} can effectively reduce the sensor-actuator delay while also providing the resiliency required in industrial automation. We also investigate the feasibility of implementing FastReact on emerging programmable data planes, discussing required features and memory requirements from the increased flexibility. 

The rest of the paper is structured as follows: In Section \ref{sec:background}, we present relevant background information. In Section \ref{sec:design}, we describe the design of \emph{FastReact}. In Section \ref{sec:evaluation}, we describe the experimental setup and our results. We conclude the paper in Section \ref{sec:conclusion} discussing future work.

\section{Background}
\label{sec:background}
In the area of SDN for industrial automation and control, there have been several works focusing on traditional SDN, where a SDN controller manages flow table entries of an OpenFlow capable switch. For example, \cite{7496525} analyses the potential for SDN in the industrial network context and identifies important research challenges such as security, reliability, real-time requirements, timing performance and Quality of Service. 
\cite{8247594}  evaluated SDN in the field of industrial automation focusing on the SDN controller. They evaluate the  performance penalty introduced by switch-controller-communication and showed that SDN is suitable for real-time traffic and has great potential for the use  in the automation industry.
\cite{7733504} proposes a system architecture for the deployment of software functionality to manufacturing resources at runtime. It is based on an integration of formal software model verification, SDN and a close interaction between a flexible runtime execution environment and its safety-critical counterpart.
\cite{7496496} finds that SDN is suitable for complex automation environments but focuses on the real-time data transmission using a Time Division Multiple Access (TDMA) mechanism combined with parallel data transmissions on physically separate links, taking application requirements into account. \cite{10.1007/978-3-319-70395-4_9}  proposes a SDN based architecture that implements response and reconfiguration capabilities in an industrial control system. \cite{jacobsson15} advocates in-network sensor data processing for robustness, reduced energy consumption and latency but modifies the sensor node architecture in order to support partial SDN functionality.

In terms of resiliency for \ac{sdn}, the centralized network monitoring allows the SDN controller to repair failed network links, providing resiliency for the network. Two main approaches are used for recovery: \emph{reactive} and \emph{proactive} link repair. When using \emph{reactive} link repair, the data plane device notifies the \ac{sdn} controller when one of their links fail. The \ac{sdn} controller uses this information to recalculate forwarding paths and install new rules in the network. While this approach is simple and memory efficient, recovery can take a significant amount of time~\cite{sharma13}. The other approach is \emph{proactive}, thus pre-installing the backup paths in the network devices, allowing for fast local failure recovery~\cite{adrichem14}. 

With the recent introduction of programmable data planes, fueled by novel switch chip architectures such as Barefoot Tofino or Cavium, programmable NetFPGAs and SmartNICs, significant limitations of SDN caused by OpenFlow \cite{tech:openflow} are removed because it allows parsing arbitrary protocol headers and stitch together much more flexible and customized switch pipelines. The introduction of language and compiler support such as P4 \cite{bosshart14} enables new use cases for SDN, because it enables the data plane to perform novel functionality. For example,  NetCache~\cite{jin17} leverages the potential of programmable data planes to perform in-network caching for key/value stores and NICached~\cite{siracusano17} enables the programmable NIC to function as a data cache. Failure recovery has been implemented using P4~OpenState through SPIDER~\cite{cascone16}. However, to the best of our knowledge, the potential of programmable data planes has not yet been applied yet to the in-network monitoring and control use case that would be required for industrial automation and control networks. A flexible and custom specification of control logic is a key problem that must be solved. 

\section{FastReact Design}
\label{sec:design}
\ngfig{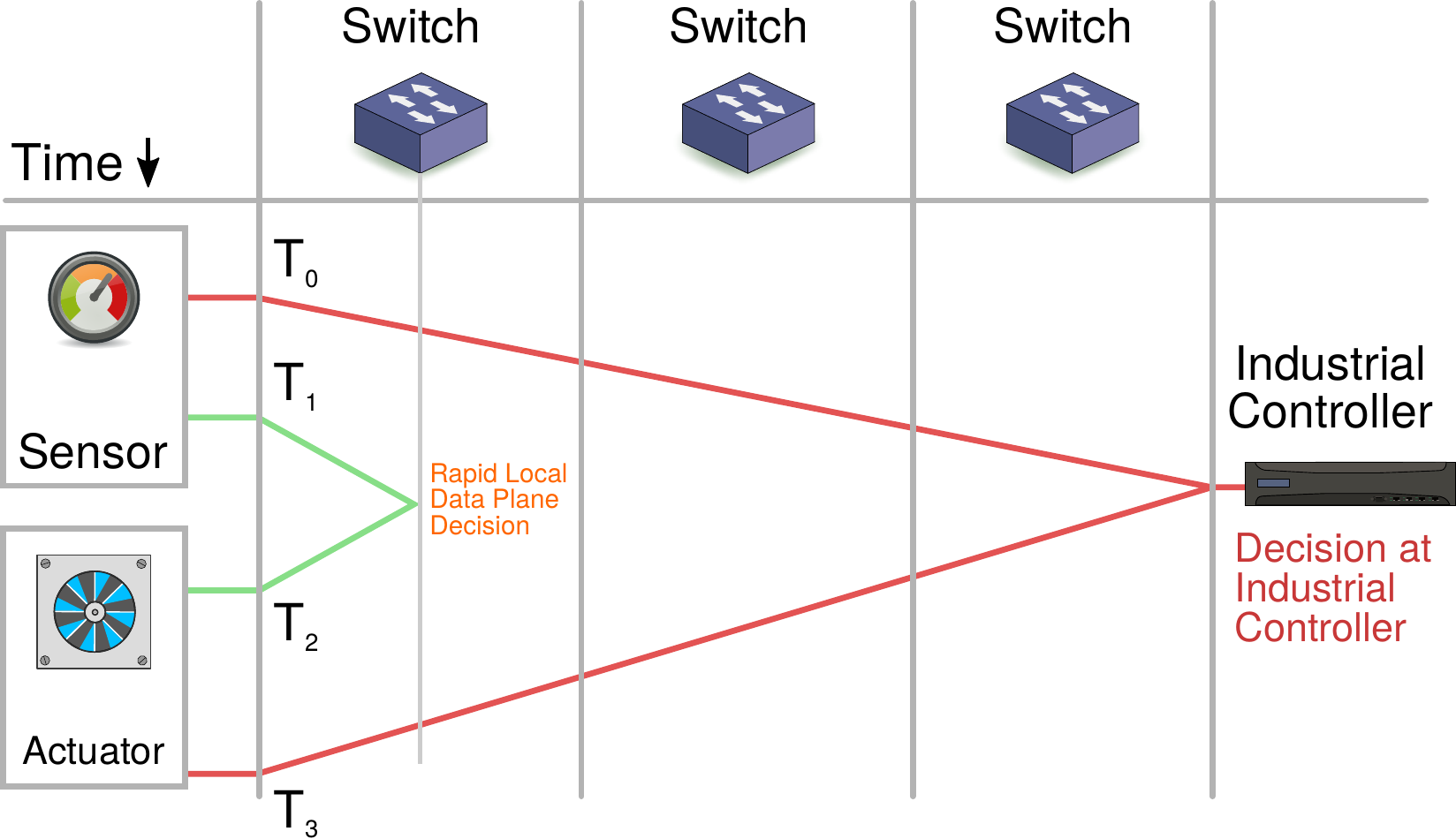}{reaction}{Example of how fast reaction could improve sensor-actuator delays in \ac{icn}.}{1\columnwidth}

\nfig{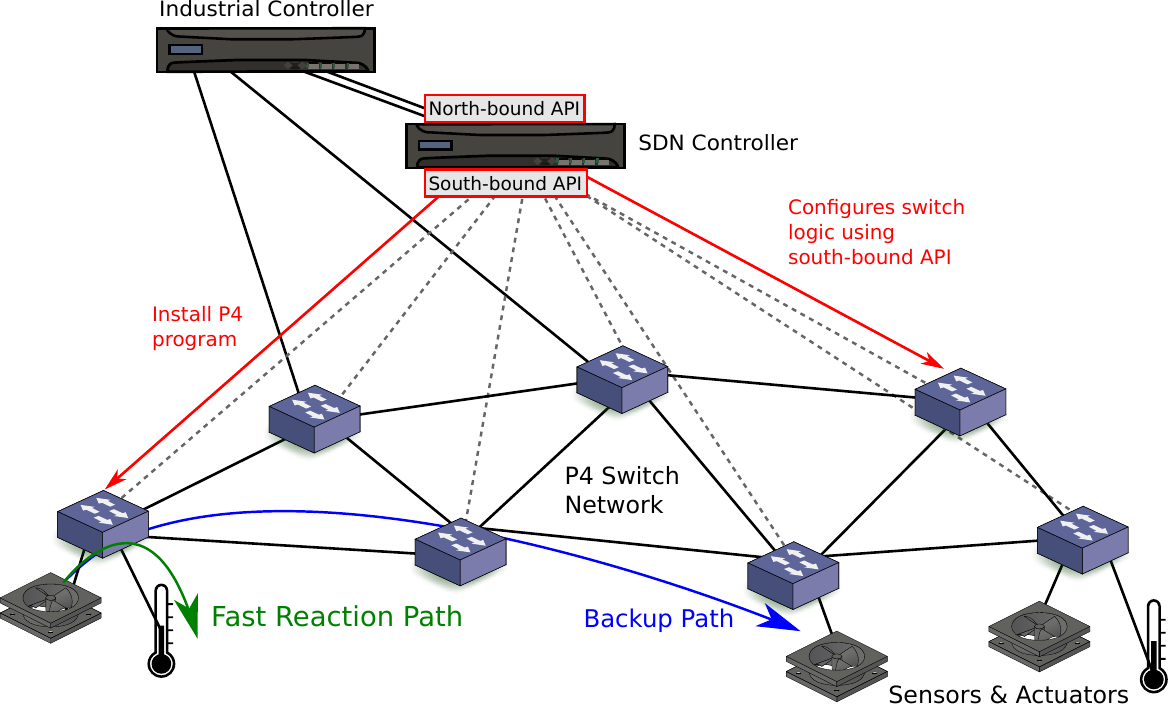}{sdn}{FastReact - Architecture Overview. }

The main idea of \emph{FastReact} is to outsource parts of the control logic of an Industrial Controller to the switch data plane, which will essentially perform in-network processing of sensor data.  \emph{FastReact} parses sensor packets using emerging programmable data plane, caches sensor values, computes lightweight operations on sensor values and, based on highly configurable logic, locally triggers actuator messages from the data plane. The microcontroller logic is configured into the \emph{FastReact} switch by a \ac{sdn} controller through populating a set of custom tables. Finally, the SDN controller exposes a North-bound API that is used by the Industrial Controller. This API allows specifying the internal processing logic of any \emph{FastReact}-enabled network switch. The SDN controller then composes processing rules using boolean expressions in normal form, which are transformed into register array entries that the SDN controller installs on the programmable switches. 

By acting locally from the data plane, the control loop between sensor and actuator can be significantly shortened, because the decisions are made closer to the sensor and/or actuators avoiding the round trip to the industrial controller (see Fig. \ref{fig:reaction}). Finally, the switch acts as a cache for sensor values and can correlate multiple sensor values to process more complex logical operations. The logic is implemented in P4 using several custom data structures implemented by registers, which allows it to be executed on P4 supported hardware. Thus, sensor packets can be processed at line rate, performing rapid, deterministic decision-making. In addition, this also increases the reliability of the network, as moving the control logic closer to the sensors and actuators reduces the potential points of failure. A high level overview on the architecture of \emph{FastReact} can be seen from Fig. \ref{fig:sdn}. 



\subsection{FastReact - Sensor Packet Parsing and Processing}
\label{ssec:history}
When a sensor packet enters the \emph{FastReact} switch, we parse the sensor \emph{ID}, extract the sensor value and store it in a time series datastore implemented by custom register arrays. We keep a rolling history of the latest $n$ values per sensor \emph{ID}, where $n$ is configurable. When inserting a new sensor value, we also compute a moving average in a separate register array using approximative arithmetic \cite{201474}. Both the historical sensor values and moving averages can be fetched by a \ac{sdn} or industrial controller either through the data plane using a \code{get}-like request, or through the control plane via the P4-generated API. In that sense, \emph{FastReact} acts similar as a key-value cache for sensor readings with the additional functionality to implement lightweight compute operations. In contrast to NetCache \cite{jin17}, which keeps only the hot items in the cache and has a cache replacement strategy, we keep a fixed number of sensor \emph{IDs} in the switch memory in order to be deterministic. Also, we argue that the Industrial Controller  knows how many sensor \emph{IDs} are deployed, so it can configure the table size during compile time of the switch. 

Based on the configured processing logic (see Section~\ref{ssec:logic}), the switch decides, if it should discard the packet, send it to the Industrial Controller or notify an actuator. The connection between sensors and actuators are configured through the match-action table, which is populated by the SDN controller as instructed by the Industrial Controller. If the conditions encoded in the logic tables are true, the actuator specified in the match-action table will be notified of the sensor value. The action will modify the sensor packet, transforming it into an actuator notification message as specified by the \ac{sdn} controller. 

An example of a match-action table can be seen in Table~\ref{tab:forward}. The table has two actions, named \code{Route} and \code{Failover}, respectively. The \code{Route} table is used to connect sensors and actuators. In this example, Sensor 1 notifies Actuator 2, while Sensor 2 notifies Actuator 5. Multiple actuators could be notified through separate actions, as P4 does not support variable number of arguments. The \code{Failover} table specifies actions the switch should take if a failure is detected. Here, for Sensor 1 the data is forwarded to Actuator 3, while for Sensor 2 the data is sent up to another switch, which, in turn, can pass the packet to a backup actuator. 

\begin{table}
\begin{small}
	\begin{tabularx}{\linewidth}{| l | l | X |}
		\hline
		{\bf Table} & {\bf Match Key} & {\bf Action} \\ \hline
		Route & sensor = 1 & forward\_mod(2 \emph{(actuator, ..)}) \\ \hline
		Route & sensor = 2 & forward\_mod(5 \emph{(actuator, ..)}) \\ \hline
		Failover & sensor = 1 & {forward\_mod(3 \emph{(bkup. actuator, ..)})} \\ \hline
		Failover & sensor = 2 & send\_up(10 \emph{(switch)}) \\ \hline
	\end{tabularx}
    \end{small}
	\caption{Example of match-action table}
	\label{tab:forward}
\end{table}

\subsection{FastReact - Decision Logic}
\def\myexpr{$(S1 < 50 \vee S2 > 25) \wedge (S3 = 10)$}
\def\mycond{$(S1 < 50)$}

\label{ssec:logic}
One of the main challenges in \emph{FastReact} is how to make the decision logic in the switch as generic as possible while being able to dynamically update it through the south-bound API. Most P4 targets expose the ability to read from and write to switch registers, which are array-like data structures stored in the switch memory. Our implementation utilizes these registers to communicate the desired control logic for each switch. In order to encode a complex logical expression in a generic way, we first transform it into its \ac{cnf}. The \ac{cnf} is given by:

\[(A \lor B \lor C...) \land (D \lor B...) \land ...\]

Where $A$, $B$, $C...$ are logical expressions in the form $(s\;{\_}\; v)$. Here $s$ represents the recorded value of a sensor, $v$ the value which it should be compared to and the placeholder $\_$ represents an operator. A simple example for a logical expression in \ac{cnf} is the following: 

\[(s_1 < 50 \lor s_2 > 25) \land (s_3 = 10)\]

Based on the processing logic provided by the Industrial Controller, the \ac{sdn} controller derives the \ac{cnf} of the logical expression and installs it on the switch, in a table format, by using switch registers. This is stored on a per-sensor basis. 

\begin{table}
\begin{small}  \begin{tabularx}{\columnwidth}{| X | X | X | X | X |}
               \hline
               {\bf \footnotesize Sensor ID} & Cond 1 & Cond 2 & Cond 3 & Cond 4 \\ \hline
               1 & 1 & 2  &   &   \\ \hline
       \end{tabularx}
    \end{small}
       \caption{Example of sensor conjunctive table for the expression \myexpr}
       \label{tab:and}
\vspace{2ex}
\begin{small}
       \begin{tabularx}{\columnwidth}{| X | X | X | X | X |}
               \hline
               {\bf Index} & Cond 1 & Cond 2 & Cond 3 & Cond 4 \\ \hline
               &\multicolumn{4}{c|}{\bf Sensor ID} \\ \hline
               1 & 1 & 2 &   &  \\ \hline
               2 & 3 &   &   &  \\ \hline
               &\multicolumn{4}{c|}{\bf Operator} \\ \hline
               1 & $<$ & $>$ &   &  \\ \hline
               2 & $=$ &   &   &  \\ \hline
               &\multicolumn{4}{c|}{\bf Value} \\ \hline
               1 & 50 & 25 &   &  \\ \hline
               2 & 10 &   &   &  \\ \hline
       \end{tabularx}
    \end{small}
       \caption{Example of sensor disjunctive tables for the expression \myexpr}
       \label{tab:or}
\end{table}

The Tables~\ref{tab:and} and Table~\ref{tab:or} shows how this data is stored in the switch. Table~\ref{tab:and} is the conjunctive table, where each row contains a sensor~ID signifying which sensor the expression should be applied to. It also contains a set of indices which point to entries in the disjunctive tables (Table~\ref{tab:or}). There are three disjunctive tables, one for sensor~IDs, one for operators and one for values, which encode the components of the comparative expression (e.g. $S1 < 50$). Each column represents a separate conditional (e.g. $A \lor B$). These tables are loaded into the switch, which transforms them into a logical expression. 

\begin{figure}
\begin{lstlisting}[label=code:p4,language=p4,basicstyle=\ttfamily\scriptsize,caption=Shortened sensor logic code.]
conj_table.read(cidx1, hdr.sensorId * CONJ_TABLE_SIZE);
conj_table.read(cidx2, hdr.sensorId * CONJ_TABLE_SIZE+1);

if (cidx1 != 0) {
	disj = false;
	disj_table_op.read(opcode, cidx1 * DISJ_TABLE_SIZE);
	disj_table_val.read(value, cidx1 * DISJ_TABLE_SIZE);
	disj_table_id.read(sensor_id, cidx1 * DISJ_TABLE_SIZE);
	sensor_latest.read(sensor_value, sensor_id);

	if (opcode == 1 && sensor_value == value) disj = true;
	if (opcode == 2 && sensor_value > value) disj = true;
	[...]

	disj_table_op.read(opcode, cidx1 * DISJ_TABLE_SIZE+1);
	[...]

	if (!disj) should_drop = true;
}

if (cidx2 != 0) {
	[...]
}
\end{lstlisting}
\end{figure}

Listing~\ref{code:p4} shows a shortened version of the P4 code for the decision logic. This code is automatically generated given parameters such as maximum number of disjunctive and conjunctive expressions, supported operators etc. As a packet arrives at the switch, the program loads the indices from the conjunctive table, corresponding to the sensor ID in the packet header. If there is logic configured (\code{cidxN != 0}), it loads the sensor value, operator, and value for the first conditional. Now, the comparative expression is performed, and if true, the program flags this (\code{disj = true}). After processing all disjunctive conditionals, if the flag is not set, the packet is dropped because one of the conjunctive terms are false. This is done for each condition in the conjunctive table. If the packet isn't dropped, the switch forwards it according to its forwarding table. 

\subsection{FastReact - Failure Recovery}
\label{ssec:failover}
In order to provide failure recovery, \emph{FastReact} monitors sensor and actuator liveness and reacts locally from the data plane should a sensor or actuator fail. In \emph{FastReact}, the switch records the timestamp of the last received packet for each switch port in a register array. Assuming that both sensor and actuators send periodic liveness messages, when no messages have been received on a particular port for a certain time, it is considered down. This port timeout interval is configured through a register array, where each entry is the timeout for a particular output port. For each actuator, a backup actuator can be specified in a failover match-action table. When a packet should be forwarded to an actuator which is considered down, the packet is instead forwarded to the backup actuator. This forwarding may be immediate (if it is reachable directly from the switch), or through another switch (as in our experiment, depicted in Figure~\ref{fig:overview}). If there are multiple backup actuators, and the \emph{FastReact} switch is directly connected to them, it will pick the first live one. If it is not directly connected, it is the responsibility of the \emph{FastReact} switch connected to the backup actuator to decide where to send the packet next. Again, it is up to the SDN controller to push the proper table entries to specify that behavior in more detail, which is outside the scope of this paper. 

\subsection{FastReact - Filtering}
\label{ssec:filtering}
To reduce the amount of traffic, \emph{FastReact} switches can filter the packets sent to the controller. This filtering can be done either using the switch decision logic programmed through the register tables, or we can only forward every \emph{n}-th packet. The filter logic is configurable on a per-sensor basis through the match-action table. For each sensor \emph{ID}, the switch keeps track of how many messages have been received and compares this count to the filtering rate, determining if the incoming packet should be forwarded or discarded. The filtering based on the decision tables is performed just like any other packet, discarding packets which do not match the logic configured through the register tables. 

\subsection{FastReact - In Network Caching}
\label{ssec:caching}
\emph{FastReact} also supports requesting stored sensor values, running averages and simple computation of the time series database through the data path using a \code{get} request. If the industrial system contains a large number of sensors, it may be beneficial to have devices request values as they are needed in order to reduce traffic. The switch keeps historical records of the latest received sensor values, along with reception timestamps. These timestamps can be used to determine the age of a certain sensor value. When the switch parses a packet in the ingress and detects a \code{get} request, the timestamp of the latest historical value for the requested sensor is compared to the current time. If the difference is smaller than a configurable tolerance, the switch forwards the request to the sensor. This request is replied to with a sensor value updated, which allows the switch to update its time series database, while also passing the message on to the original requester. To instruct the switch what kind of data it should return (e.g. most recent value, moving average), an OpCode is provided in the \code{get} request with the sensor \emph{ID}.

\subsection{FastReact - Feasibility in P4}
\label{ssec:feasibility}
Our P4 implementation requires support for basic P4 primitives defined by the P4\_16 standard~\cite{budiu17}. In addition, it requires support for registers reading, register writing, IP and UDP checksum calculation and stateless header parsing. Support for ingress timestamps is also required. 

\textbf{Sensor Dependency Table:} The switch memory requirements depend on the logic table sizes, the number of sensors supported and the history size. The size of the conjunctive table is $S_{count} C_{cols} \lceil\log_2{D_{rows}}\rceil$ bits. 
Here $S_{count}$ is the number of sensors, $C_{cols}$ is the maximum number of conjunctive conditionals and $D_{rows}$ is the maximum number of rows in the disjunctive table. 
The size of the disjunctive table is $(D_{rows} D_{cols}) (3 + Sz_{sen} + \lceil\log_2{S_{count}}\rceil)$ bits. 
$D_{rows}$ is the maximum number of rows in the disjunctive table, $D_{cols}$ is the maximum number of disjunctive conditionals, $Sz_{sen}$ is the size of the sensor value data type and $S_{count}$ is the number of sensors. 
Calculating an appropriate disjunctive table size requires some consideration, because it represents the maximum number of conditional expressions in the logic table. 
For example, using 16-bit sensor values, 5000 sensors, a conjunctive table size of 25.000 and a maximum of 5 disjunctive and 5 conjunctive conditionals, the dependency table memory requirement amounts to around~\SI{4.4}{MB}.

\textbf{Sensor Time Series Database:} In order to store historical sensor values and their respective timestamps, the memory requirements are: $(H_{count} + 1) (Sz_{ts} + Sz_{sen})  S_{count}$ bits of storage. $H_{count}$ is the number of values to store per sensor \emph{ID}, one additional entry is used for storing and updating the moving average. $Sz_{ts}$ and $Sz_{sen}$ are the timestamp and sensor value data type size. Finally, $S_{count}$ is the number of sensor \emph{ID}s. 
In addition, to store the round robin indexes that determine, which is the sensor value received last, we additionally require: $\lceil\log_2{H_{count}\rceil S_{count}}$ bits of storage. 
For 5000 sensors \emph{ID}s and 100 historical values per sensor, we need approximately~\SI{32}{MB} assuming 16-bit sensor value data type and 48-bit timestamps. 

\textbf{Other Memory Requirements: } The failure recovery requires storing one timestamp for each switch port. The filtering requires storing a counter for each sensor \emph{ID}. For 5000 sensors and 24 switch ports, this will amount to around~\SI{80}{KB} assuming 16-bit sensor counters and 48-bit timestamps. Furthermore, some settings are stored in registers, but require minimal space (less than \SI{1}{KB}). 

\subsection{FastReact - Implementation}
\label{ssec:implementation}
\nfig{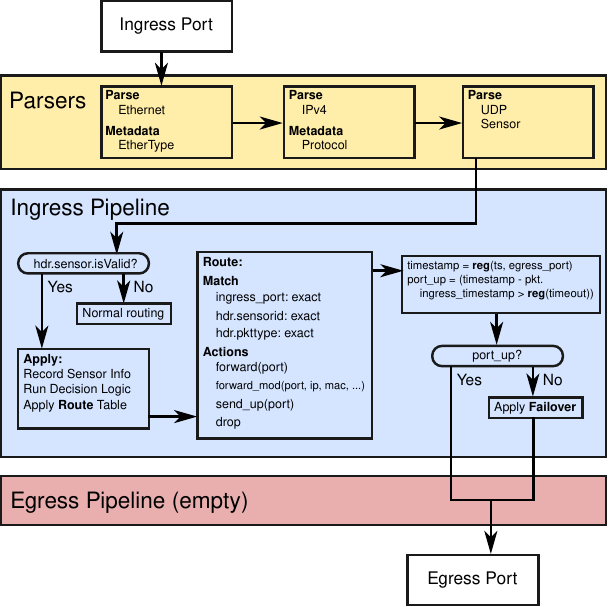}{p4code}{Overview of our implementation of \emph{FastReact} in P4. }

In order to evaluate the feasibility of our design, we implemented a prototype in P4 using the \code{v1model} architecture\footnote{The P4 code will be available at \url{https://github.com/jonavest/P4-FastReact}}. As can be seen from Fig.~\ref{fig:p4code}, packets are parsed, determining if they are sensor messages, \emph{get} requests (not shown for simplicity) or regular network traffic. If a sensor packet is detected, the sensor value is extracted and  recorded in the time series registers. Then, the decision logic is applied, which determines if the packet should be dropped, sent to another switch or an actuator should be notified. If an actuator is notified, the \code{route} match-action table instructs the switch, based on the sensor \emph{ID}, ingress port and packet type what action should be performed. The \code{forward(port)} action simply forwards the packet normally and the \code{forward\_mod(port, ip, mac)} action sends a notification to an actuator. 

Then, the ingress timestamp of the packet is compared to the latest timestamp of the output port. If the difference is greater than the configured tolerance, the \code{failover} table is applied. The \code{failover} table  can only result in the \code{send\_up(port)} action. The \code{send\_up(port)} changes the packet type to indicate that it is using a backup route and sends it to the specified port. This indication is later used by other switches to help determine the appropriate backup destination. Finally, checksums are recalculated and the packet is moved to the egress pipeline. 

Filtering is performed after the decision logic, where a filtered packet counter is incremented, and compared to the filter rate, determining if the packet should be filtered (dropped) or forwarded. Cache timestamps are updated when the sensor value is recorded, while \code{get} requests are processed in the beginning of the pipeline. These \code{get} requests are handled separately from the sensor logic. When a \code{get} request enters the switch, the timestamp of the latest historical sensor entry is compared to the ingress packet timestamp. If the difference is smaller than a configurable tolerance, a new sensor value is requested from the sensor. Otherwise, the operational code is extracted and the operation on the time series registers are calculated and returned from the cache. 


\section{Evaluation}
\label{sec:evaluation}
In this section, we evaluate the effectiveness of FastReact by implementing it in P4 and performing several experiments using the CORE network emulator~\cite{ahrenholz08}. CORE uses Linux Network Namespaces to create virtual hosts and links. These are run using the normal Linux networking stack, which allows integration of the P4 behavioral model\footnote{Available at \url{https://github.com/p4lang/behavioral-model}}. We use the modified CORE version\footnote{Available at \url{https://github.com/tohojo/core}} which uses \code{netem} instead of \code{tbf} for rate management, which allows to tune the queue length. 
\subsection{Experimental Setup}
\label{sec:experiment}

We set up a testbed (see Fig. \ref{fig:overview}) for validating the basic functionality and performance characteristics. The industrial controller is connected to the switch $C1$, which in turn is connected to the switch $B1$. $B1$ is connected both to switches $A1$ and $A2$. Both switches have one sensor and one actuator connected, named \emph{Actuator 1} and \emph{Sensor 1} for $A1$, and \emph{Actuator 2} and \emph{Sensor 2} for $A2$ respectively. 
Each link is configured with a latency of \SI{1}{\ms}, and a throughput of \SI{1}{Gbps}. Therefore, the delay between \emph{Sensor 1} and \emph{Actuator 1} is \SI{2}{\ms}, while the delay between \emph{Sensor 1} and \emph{Actuator 2} is \SI{4}{\ms}. For the path Sensor $\rightarrow$ Controller $\rightarrow$ Actuator, the total delay is \SI{8}{\ms}. 

All experiments are run on a Dell Optiplex~9020 (4-core i7~4790~3.6GHz, \SI{8}{GB}~RAM), running version 4.1.39-rt47 of the Linux kernel. The Linux kernel has been patched using the \code{PREEMPT\_RT} patch, running with the \code{PREEMPT\_RT\_FULL} model, in order to reduce latency and jitter caused by process scheduling, while improving timestamp accuracy. 

\ngfig{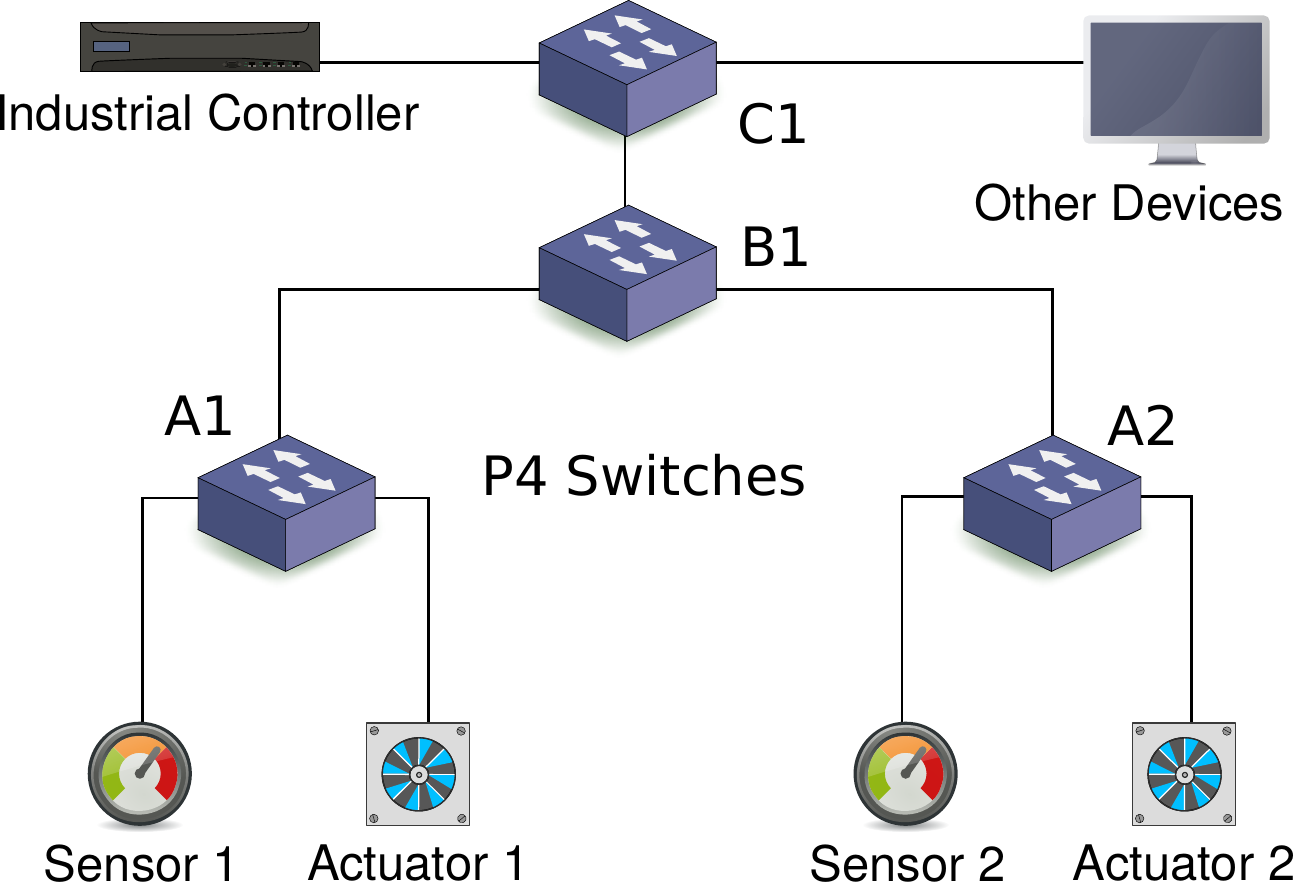}{overview}{Emulated experiment testbed running in CORE. }{.3\textwidth}

\subsection{Results}
\label{sec:results}
\subsubsection{Baseline}
\nfig{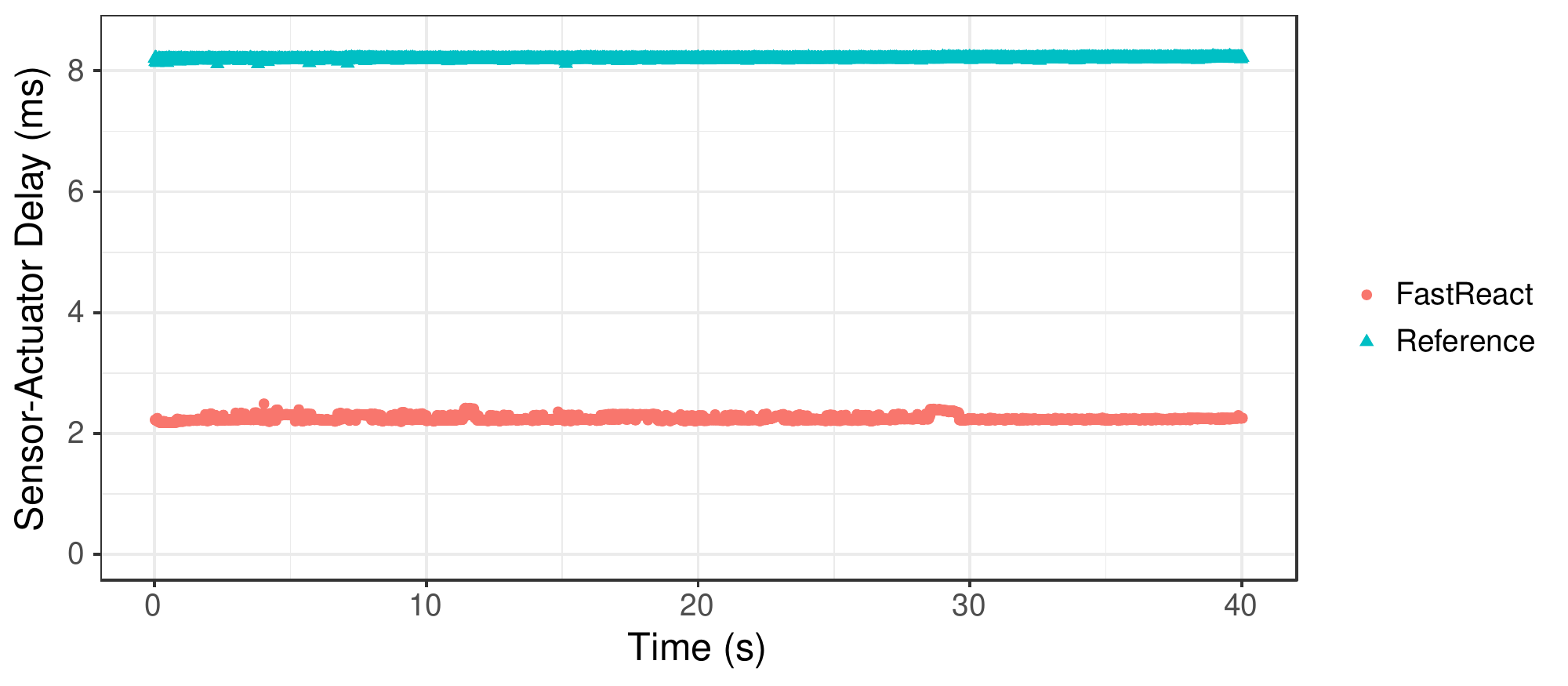}{ref}{Reference experiment using a single  sensor and actuator. The ``Reference'' test uses normal controller processing, while ``Fast Reaction'' uses \emph{FastReact}. }

First, we performed a reference experiment using traditional switching, which is not aware of the sensor format and can not react to changes in the sensor readings. All sensor messages are sent to the industrial controller which then sends a reply to the appropriate actuator. In this experiment, \emph{Sensor 1} sends continuous sensor messages to the industrial controller over \SI{40}{\s} with an interval of \SI{10}{\ms}. The industrial controller will receive these messages and passes an appropriate action to \emph{Actuator 1}. Fig. \ref{fig:ref} shows the results of the experiment. In the figure, the x-axis represents the time when \emph{Sensor 1} sent the packet, and the y-axis represents the latency between \emph{Sensor 1} and \emph{Actuator 1}. We compare the delay with our approach \emph{FastReact}, where the P4 enabled switch parses sensor packets in the data plane, checks the control actions and directly sends commands from the switch to the actuator. As we can see from our experimental data, the sensor-actuator delay averages to \SI{8.201}{\ms} for the case where the Industrial Controller decides the action  compared to an average delay of \SI{2.249}{\ms} for \emph{FastReact}. The significant reduction in latency is because the sensor messages only pass through the switch, and we avoid the costly communication to reach the industrial controller. 

\subsubsection{Failover}
\nfig{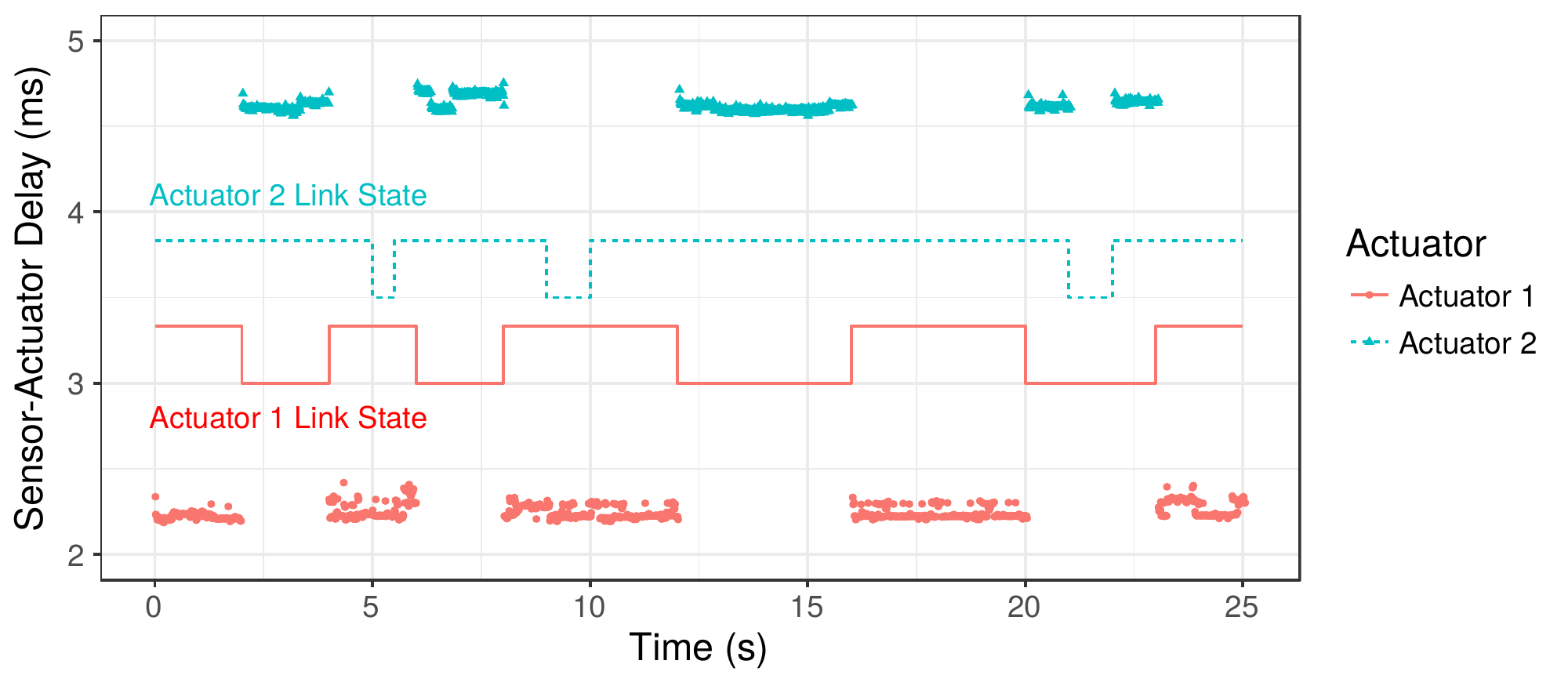}{multifail}{Sensor-Actuator delay when the link between the actuator and their corresponding switch go up and down. }

In order to evaluate the failure recovery behavior of \emph{FastReact}, we performed a second set of experiments. First, we evaluated how resilient the network is to a single link failure. In this experiment, \emph{Sensor 1} sends messages at a \SI{10}{\ms} interval and the port timeout is set to \SI{30}{\ms}. If the switch does not receive any packets for \SI{30}{\ms} on a port, that port is considered to be down, and the message will be forwarded up the network, to switch $B1$. $B1$ sends the message down to switch $A2$ which sends it down to \emph{Actuator 2}. Fig. \ref{fig:multifail} shows sensor-actuator delays over time, and the lines show the link states of \emph{Actuator 1} and \emph{Actuator 2} respectively. As we can see, when the link between \emph{Actuator 1} and $A1$ or the node \emph{Actuator 1} fails, \emph{FastReact} immediately reacts by sending the sensor values through $B1$ and $A2$ to \emph{Actuator 2}, which causes an increase in the latency from an average of \SI{2.244}{\ms} to \SI{4.624}{\ms}. This increase is due to the longer path between the sensor and the actuator. It is worth pointing out that if both links between \emph{Actuator 1} and $A1$ and between \emph{Actuator 2} and $A2$ fail, no data is received, because no actuator is reachable. The number of packets lost during the local repair in this scenario is between 2 and 3. 

One important aspect of how fast the switch can recover from a link failure is determined by the packet sending interval configured in the actuator and the port timeout configured in the switch. The shorter the sending interval, and the shorter port timeout, the faster faults can be detected. However, using a short sending interval increases the amount of traffic on the link, and using a short port timeout increases the probability for false positives (due to unexpected packet loss or jitter). Fig. \ref{fig:failover} shows the sensor-actuator delay with varying actuator sending intervals and port timeouts. In this experiment, packets are sent between \emph{Sensor 1} and \emph{Actuator 1}, as the link between \emph{Actuator 1} and $A1$ is taken down. The measurement points in the figure are the difference between the last received packet for \emph{Actuator 1} and the first received packet for \emph{Actuator 2}. From the figure, we can see that the recovery time is highly impacted by both sending interval and port timeout. For short port timeouts of \SI{10}{\ms} in combination with a short sending interval of \SI{5}{\ms}, the mean sensor-actuator delay (during failure) is \SI{14.45}{\ms}. 
The missing measurement points are in the cases when the sending rate is slower than the port timeout, causing both ports to be in a down state, and no packets are being transmitted.

\nfig{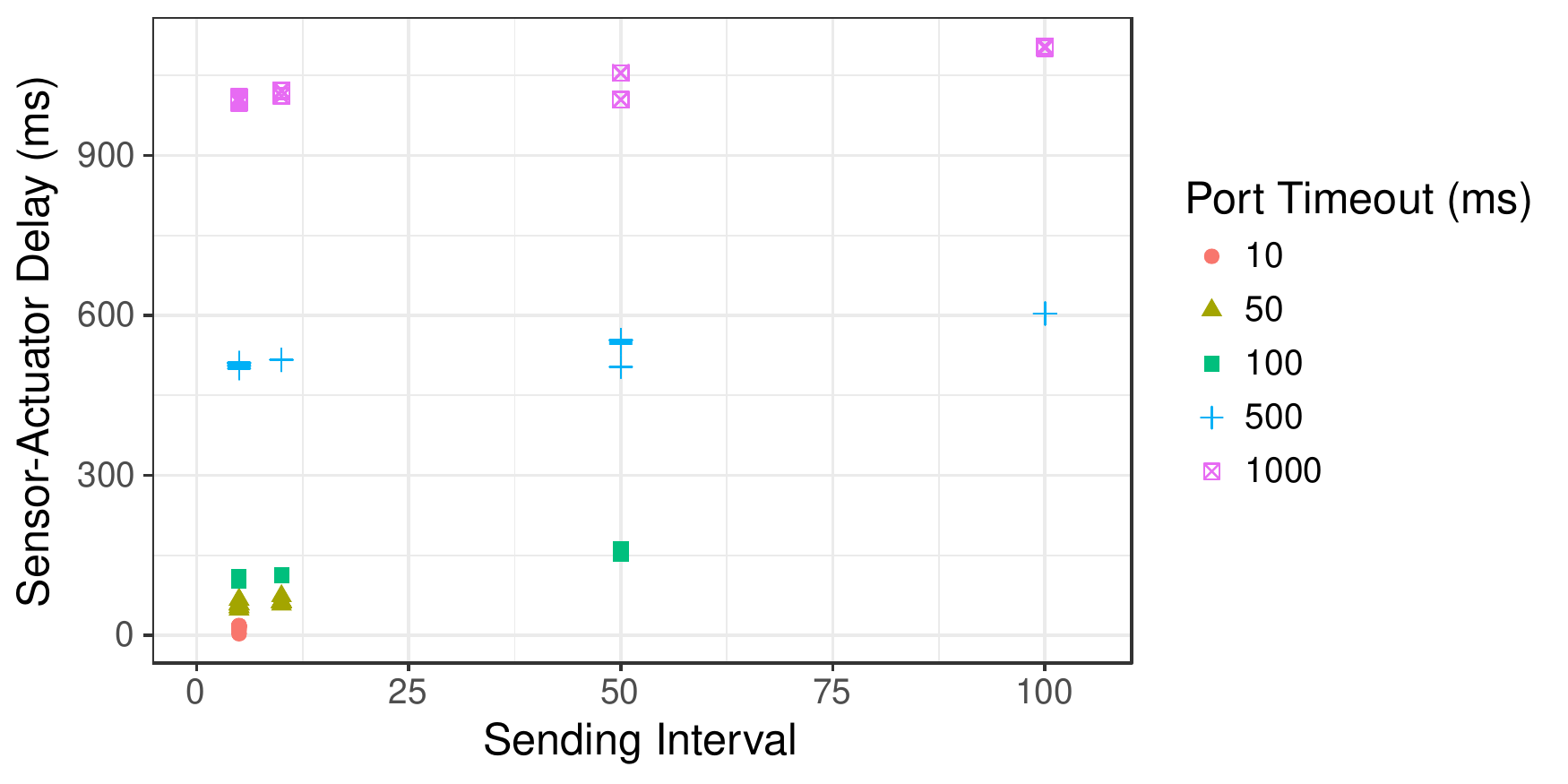}{failover}{Sensor-Actuator delays in case of a failure with varying actuator sending intervals and port timeouts. }
\note{J: Confirm these results especially at the lower end. }

\subsubsection{FastReact Logic Processing}
\emph{FastReact} is capable to correlate multiple sensor values in order to determine proper actions as given by the control logic in the data plane. This section shows results from experiments focusing on the impact of the switch processing logic on delay. 

\nfig{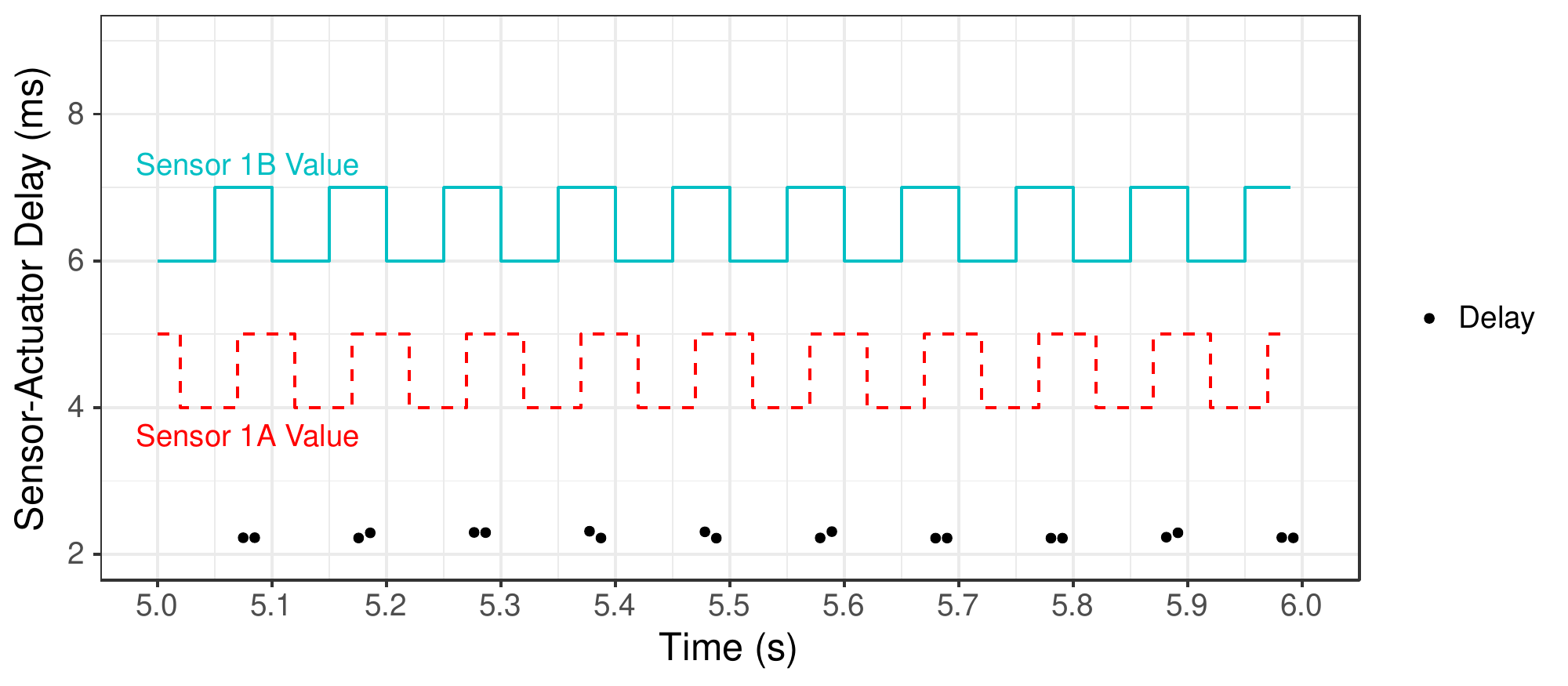}{bindep}{Sensor-Actuator delays with binary sensor values. }

Fig. \ref{fig:bindep} shows an experiment where we have used two sensors which take binary values. A message should be sent to the actuator only when both sensors are sending the value $1$. The two lines represent the sensor value at the current time, while the dots represent the arrival time of messages at the actuator. In this test, two sensors are configured which alternate between sending the values $0$ and $1$ every \SI{0.5}{\s}, with one of the sensors shifted its starting time slightly. The test was run for \SI{20}{\s}, however, the figure shows only the results between \SI{5}{\s} and \SI{6}{\s}. 
From the figure, we can see that the actuator only receives messages when both sensors observe a value of $1$. This behaviour is of course fully configurable depending on the application.

\nfig{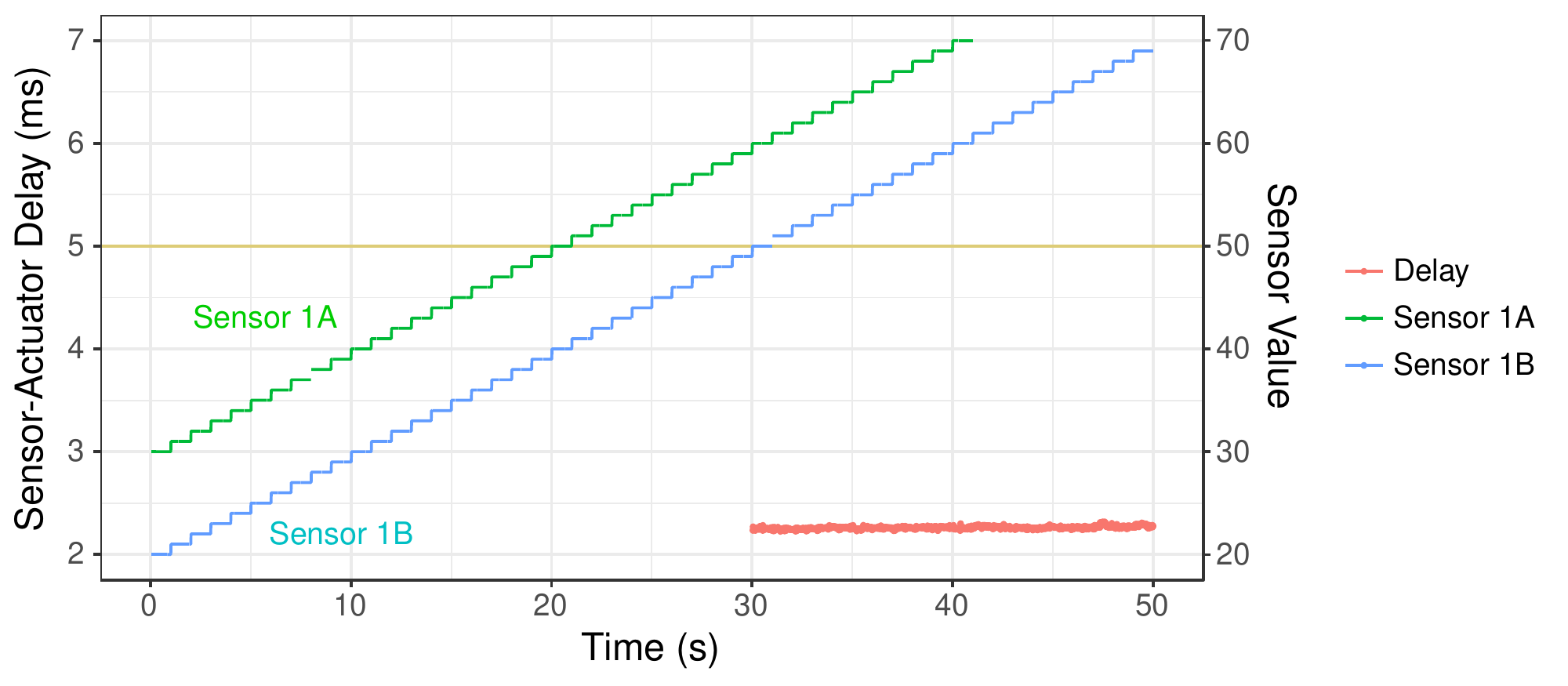}{double}{Experiment using two sensors with stateful switching. }

In the next experiment, we have two sensors, \emph{Sensor 1A} and \emph{Sensor 1B}, which report integer values. The \emph{FastReact} logic is configured so that when the reported value exceeds $50$ for both sensors we want to notify the actuator, i.e. $(A \geq 50) \wedge (B \geq 50)$. Because we record the historical values for each sensor in the data plane in \emph{FastReact}, we compare the received sensor value and correlate it with the previously received value of the other sensor. Fig. \ref{fig:double} shows two dashed lines representing the value of the two sensors, and a solid line representing the delay between sensor and actuator. The sensor values increase  by $1$ unit every second. The sensors are configured to start at $20$ and $30$ respectively. 
As we can see from the figure, the switch is able to keep the state of both sensors in memory, and wait for both sensors to reach the configured value before it triggers the fast reaction and notifies the actuator.

\note{J: When redoing the graphs, please redo two values so that the sensor names are correct}

\nfig{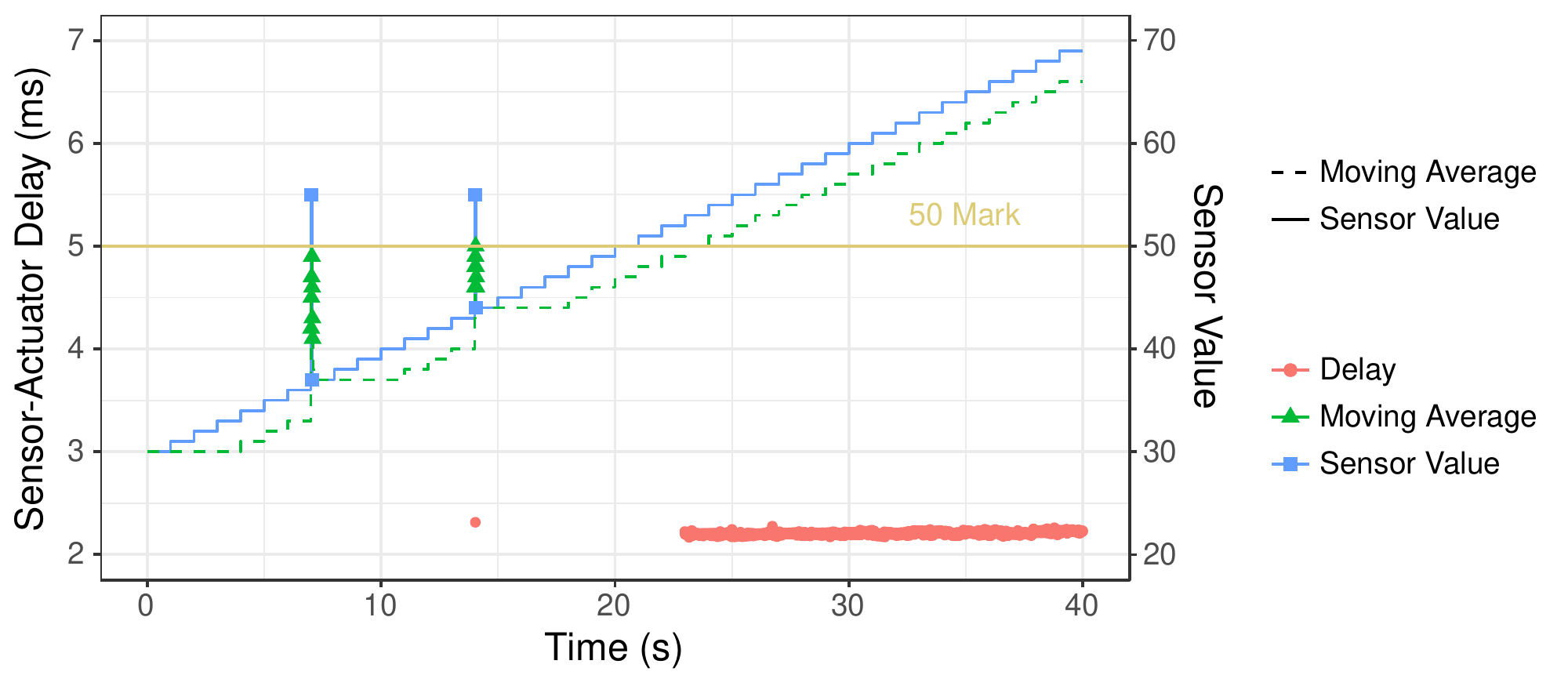}{spikes}{Experiment using single sensor and actuator. Sensor values suffering from spikes. Using moving average values instead of latest. Points are drawn on the figure lines during the spikes for readability. }

In order to demonstrate the moving average functionality, we constructed an additional scenario where the sensor values occasionally spike. If we were just looking at the last sensor value, we would notify the actuator intermittently, causing maybe an unnecessary activation of the actuator leading to state oscillation. The spikes occur with an interval of \SI{7}{\s} and changes the sensor value to $55$ for \SI{50}{\ms}. As we can see from 
Fig. \ref{fig:spikes}, two spikes occur at $t=7$ and $t=14$. The first does not trigger an alert to the actuator, because the moving average evens out the matched value. However, the later spike (at $t=14$) triggers an immediate fast reaction and sends a message directly to the actuator, because the moving average will increase above $50$.

\nfig{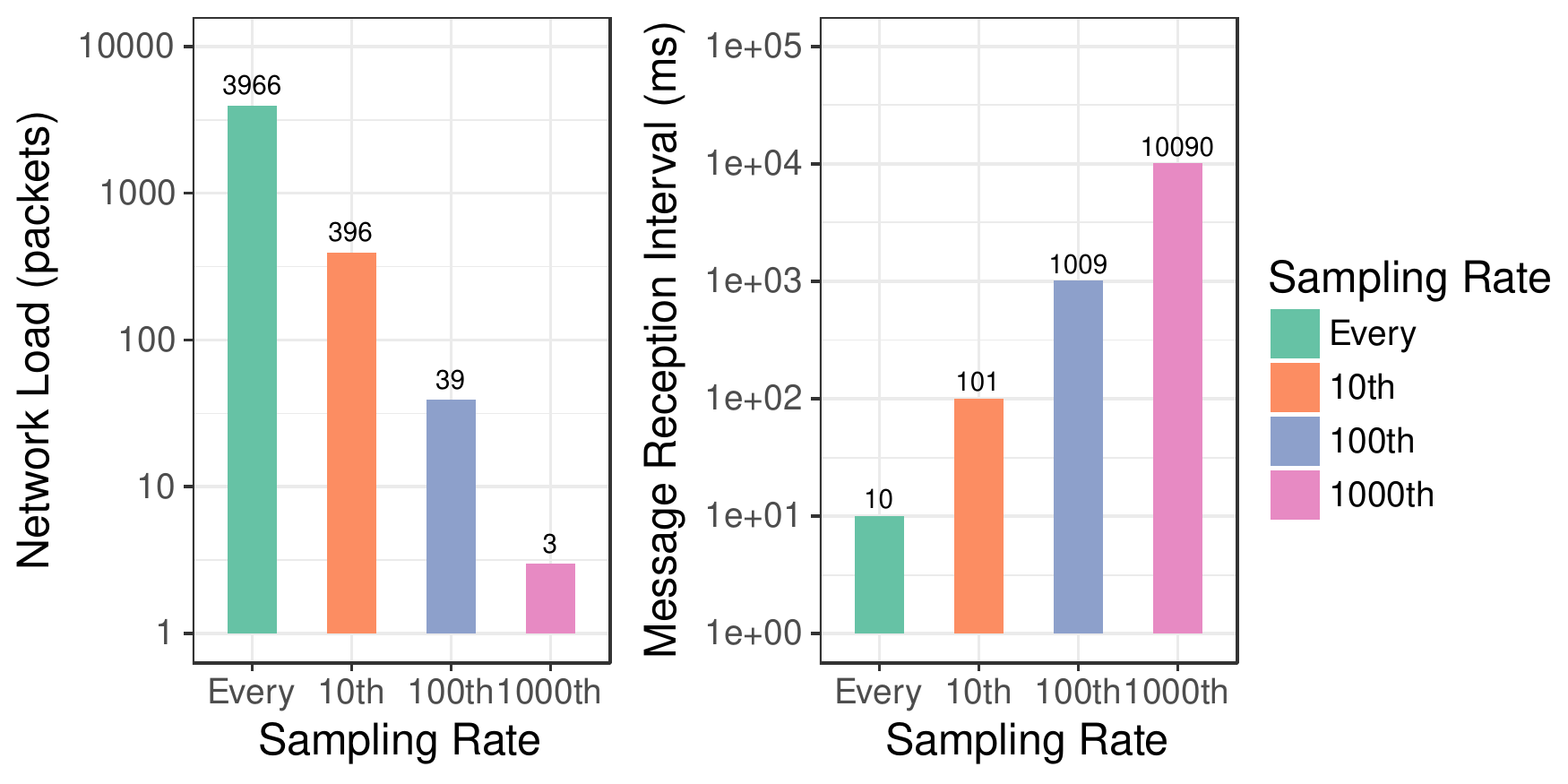}{filter}{Experiment filtering packets at various sampling rates.}

Fig. \ref{fig:filter} shows the effect of filtering sensor messages at the switch. In this experiment, messages are always sent from the sensor to the industrial controller, and then forwarded to the actuator. No fast reaction is employed, however, packet filtering is enabled with various sampling rates. The sampling rate can be seen in the x-axis of the figure while the y-axis is the network load (in packets) and the time between messages received at the controller. The sensor-actuator delay show the increase in delay due to filtering messages. Decreasing the sampling rate reduces the load on the network, while increasing the potential sensor-actuator delay.

\note{J: Check notation}

\nfig{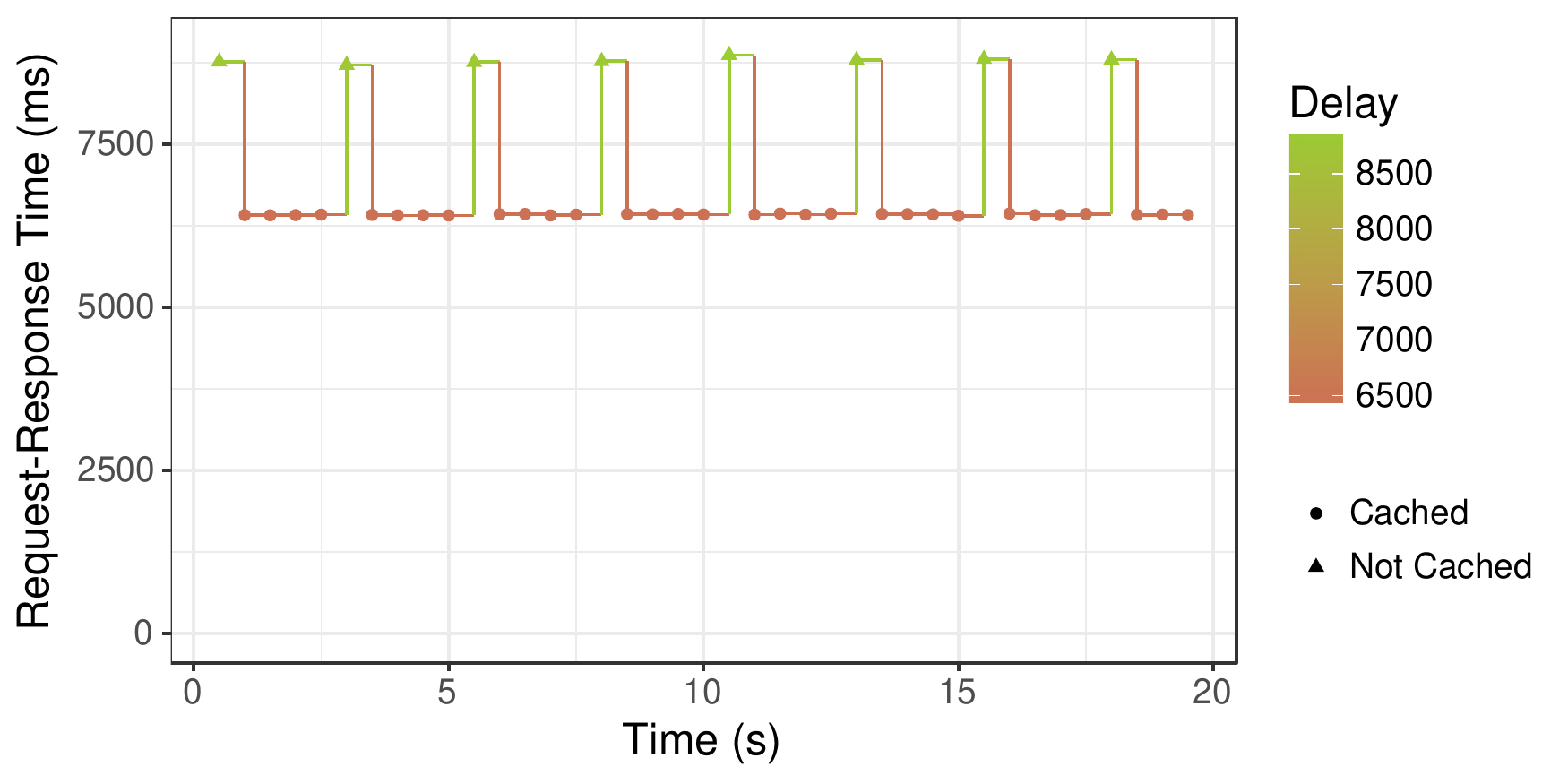}{reqlong}{Time between request and response for \code{get} requests sent from the industrial controller to a sensor node. }

Finally, Fig. \ref{fig:reqlong} shows the industrial controller continuously sending \code{get} requests to the sensor node. As the cache timeout is set to \SI{5}{\s}, the first request is transmitted through $A1$ to the sensor, while in the 4 subsequent requests the value has been cached by \emph{FastReact}. As such the cached sensor value is immediately returned by the switch. The y-axis shows the time between the controller request was sent until the response was received. 
The ``Not Cached'' dots shows the request time when there is no cache entry, while the ``Cached'' dots shows the request time when the sensor value is in the switch cache. As we can see from the figure, the average delay between request and response is \SI{8.73}{\ms} if the value has not been cached, and \SI{6.38} {\ms} if the value has been cached by the switch.

\section{Conclusion}
\label{sec:conclusion}
In this paper, we have introduced \emph{FastReact}, a in-network monitoring, control and caching system for industrial automation. Based on highly configurable logic expressed in boolean algebra, the SDN controller tailors the control logic of the switch data plane as instructed by the industrial controller. \emph{FastReact} switches perform lightweight computational operations and local control actions, while being resilient to link or sensor/actuator node failures. We have presented a high-level design of the system, and discussed the feasibility of implementing it on emerging programmable  switch platforms. We discuss scalability issues in relation to implementation and platform  constraints as given by the P4 language. Finally, we have implemented and evaluated basic functionality and performance characteristics, showing that \emph{FastReact} can significantly reduce the sensor/actuator delay, while still providing high resiliency to network failures. 

As for future work, we intend to evaluate the performance impact of \emph{FastReact} on the data plane performance using NetFPGA and programmable network cards. In addition we intend to focus on SDN controller interaction with Industrial controller for a more flexible design of the north-bound API.

\section*{Acknowledgement}

Parts of this work has been funded by the Knowledge Foundation of Sweden through the project READY.

\def\bibfont{\footnotesize}
\printbibliography
\end{document}